\documentclass[12pt]{iopart}

\begin{document}

\title{On Lame's equation of a particular kind}



\author{Pavel Ivanov}

\address{Theoretical Astrophysics Center, 
Juliane Maries Vej 30, Dk-2100 Copenhagen, 
Denmark and Astro Space Center of P. N. Lebedev Physical Institute, 84/32 
Profsoyuznaya Street, Moscow, 117810, Russia}


\begin{abstract}

It is shown that Lame's equation ${d^{2}\over d z^{2}}X+\kappa^{2}
cn^{2}(z, {1\over \sqrt 2})X=0$
can be reduced to hyper-geometric equation. 
The characteristic exponents of this equation 
are expressed in terms of elementary
functions of the parameter $\kappa$. 
An analytical condition for parametric amplification is obtained. 

\end{abstract}

\maketitle

\section{Introduction}

Lame's equation
$${d^2\over dz^2}X+(R_1+R_2 sn^2(z, K))X=0, \eqno 1$$
where $sn (z, K)$ is the standard Jacobian elliptic sine function of modulus $K$
\footnote{The elliptic sine function $sn(u,K)$ and the elliptic cosine function $cn(u,K)$ are
defined with help of the elliptic integral 
$u=\int^{\phi}_{0}{d\alpha \over \sqrt {1-K^{2}\sin^{2}(\alpha)}}$. Then, $sn(u,K)=\sin(\phi)$
and $cn(u,K)=\cos (\phi)$.}, and
$R_1$, $R_2$ are numerical parameters, is common in many branches of mathematical physics.

Attention to this equation has been called recently by the fact that it plays an
important role in certain problems of particle physics and cosmology of the Very Early
Universe. In particular, it was shown [1] that in the theory of two fields
$\phi$ and $\chi$ with potential
$V(\phi, \chi)={\lambda \phi^{4}\over 4}+{g^{2}\phi^{2}\chi^{2}\over 2}$ 
the equation of motion of the field $\chi$ in the Heisenberg representation
can be reduced to Lame's equation under certain assumptions.  
More concretely, for the Minkowski space-time 
in the linear approximation for the field $\chi$, it is found that after expansion of
this field over the eigenfunctions of the Laplace operator, the equation of motion for
an eigenfunction corresponding to a particular wavenumber has 
the form 
$${d^{2}\over dz^{2}}X +(k^{2}+{g^{2}\over \lambda}cn^{2}(z, {1\over \sqrt{2}}))X=0,
\eqno 2$$ 
where $cn(z, {1\over \sqrt 2})$ is the Jacobian elliptic cosine function of
modulus $K={1\over \sqrt 2}$, 
the parameter $k$ is proportional to the wavenumber, 
and the ``time'' $z$ is proportional to the ordinary Minkowski time coordinate. In fact,
the same equation is valid in the expanding Universe  
in a certain regime provided that the eigenfunctions are rescaled in a proper way, and the
coordinate $z$ plays the role of so-called conformal time. 

Along the real axis, $cn(z, {1\over \sqrt 2})$ is a periodic function with 
period $T=4{\bf K}({1\over \sqrt 2})={\Gamma^{2}(1/4)\over \sqrt \pi}\approx 7.416$,
where ${\bf K}(K)$ is the complete elliptic integral of modulus $K$ of first type 
and $\Gamma(x)$ is the gamma function.   It follows from  general theorem that 
the equation (2)
must contain solutions in the form $X_{1,2}=e^{\mu_{1,2}z}P(z)$, where $P(z)$ is 
a periodic function of the period $T$, and the coefficients $\mu_{1,2}$ are called 
characteristic exponents. If one of these coefficients is real and positive, the 
corresponding solution describes an exponential growth of the amplitude of the
eigenfunction $X$. In modern theories of matter creation in the Universe [2-4]
this growth is interpreted as a production of ``particles'' of the field $\chi$. The
rate of production of the  ``particles'' is determined by the values of characteristic exponents,
and therefore the calculation of these exponents is very important for such theories.
Usually the calculation of characteristic exponents is performed by numerical means,
and only a few cases are known with analytical solutions. In particular,
the characteristic exponents have been calculated analytically 
in the paper [1] for the case ${g^{2}\over \lambda}={n(n+1)\over 2}$ (n is an integer). 

In this note we would like to point out that the special case of the equation (2) with
$k=0$: 
$${d^{2}\over dz^{2}}X +\kappa^{2}cn^{2}(z, {1\over \sqrt{2}})X=0,
\eqno 3$$
($\kappa ={g^{2}\over \lambda}$)
can be reduced to the hyper-geometric equation. This allows for exact calculation of 
the characteristic
exponents for that important case
\footnote{Which corresponds to a long wave approximation from viewpoint 
of particle physics and cosmology.},   
representing them in a remarkably simple form (see eqs. 21, 25 below).

\section{Reduction of the equation to the hyper-geometric equation and the characteristic
exponents}

Let us consider the equation of form (3), and make the following change of the
independent variable:
$$y=cn^4(z,{1\over \sqrt 2}). \eqno 4$$
Then, we use the well known relation 
$${d \over d z} cn(z, K) =\sqrt {(1-cn^{2}(z, K))(K^{'2}+K^{2}cn^{2}(z, K))},\eqno 5 $$
where $K^{'}=\sqrt{(1-K^{2})}$. Taking into account that in our case 
$K^{'}=K={1\over \sqrt 2}$, we obtain
$${d\over dz}=\pm 2\sqrt 2 y^{3/4}\sqrt{(1-y)}{d\over dy}, \eqno 6$$
where the $+$ sign ($-$ sign) should be taken if $y$ increases (decreases)
with $z$. Also, we have
$${d^{2}\over dz^{2}}=8y^{1/2}\lbrace y(1-y){d^{2}\over dy^{2}}+
({3\over 4}-{5\over 4}y){d\over dy}\rbrace . \eqno 7  $$ 
Substituting the last operator into equation (3), we see that the factor
$y^{1/2}$ cancels and that this equation
is transformed to the standard hyper-geometric form:
$$y(1-y){d^2\over dy^{2}}X+(\gamma-(\alpha+\beta+1)y){d\over dy}X-\alpha \beta X=0,
\eqno 8$$
where $\gamma={3\over 4}$, and $\alpha, \beta ={1\over 8}(1\pm \sqrt{(1+8\kappa^2)})$.
The general solution to eq. (8) can be written in terms of two elementary 
solutions valid in the vicinity of the singular point $y=0$:
$$X=c_1 \phi_1 +c_2 \phi_2, \eqno 9$$
where
$$\phi_{1}=F(\alpha, \beta, \gamma; y) \eqno 10$$
is the Gaussian hyper-geometric function, and
$$\phi_2=y^{1-\gamma}F(\alpha+1-\gamma, \beta +1-\gamma, 2-\gamma; y). \eqno 11$$
Alternatively, the solution to eq (8) can be written in terms of two elementary solutions
valid in the vicinity of the singular point $y=1$:
$$X=c_3\phi_3 +c_4 \phi_4, \eqno 12$$
where  
$$\phi_{3}=F(\alpha, \beta, \alpha +\beta+1-\gamma ; 1-y), \eqno 13$$
and
$$\phi_{4}=(1-y)^{\gamma -\alpha -\beta}
F(\gamma -\alpha, \gamma -\beta, \gamma+1-\alpha-\beta ; 1-y). \eqno 14$$
The pairs of solutions $\phi_1$, $\phi_2$ and $\phi_3$, $\phi_4$ are connected by the
well known relations (e. g. [5])
$$\phi_1=A\phi_3+B\phi_4, \quad \phi_2=C\phi_3 +D\phi_4, \eqno 15$$
$$A={\Gamma(\gamma)\Gamma (\gamma -\delta)\over 
\Gamma (\gamma -\alpha) \Gamma (\gamma -\beta)}, \quad
B={\Gamma(\gamma)\Gamma (\delta - \gamma)\over 
\Gamma (\alpha) \Gamma (\beta)} \eqno 16$$
$$C={\Gamma(2-\gamma)\Gamma (\gamma -\delta)\over 
\Gamma (1 -\alpha) \Gamma (1 -\beta)}, \quad
D={\Gamma(2-\gamma)\Gamma (\delta - \gamma)\over 
\Gamma (\alpha+1-\gamma) \Gamma (\beta+1-\gamma)}, \eqno 17$$
where $\Gamma(x)$ is the Gamma function, and $\delta=\alpha+\beta$.

Let us point out that the transformation (4) is singular at the points 
where the function y is equal
to zero or unity (${d y\over dz} \rightarrow 0$ when
$y\rightarrow 0, 1$). 
A simple analysis shows that the coefficient $c_{2}$ must change its sign at the
points $z_{i}$ defined by $y(z_{i})=0$, and the coefficient
$c_{4}$ must change its sign at the points $z_{j}$ satisfying $y(z_{j})=1$.
To show that, let us consider the behavior of the function $X$ near the points
$z_{i}$. Near these points, we can approximately write
$$X\approx c_{1}+c_{2}y^{1/4}, \eqno 18 $$
and taking into account equation (6), we have
$${dX\over dz}\approx -{c_{2}\over \sqrt 2}, \eqno 19$$
provided ${dy\over dz} < 0$, and 
$${dX\over dz}\approx {c_{2}\over \sqrt 2}, \eqno 20$$
provided ${dy\over dz} > 0$. The function $X$ and its derivative with respect
to ``time'' $z$ must be continuous functions of $z$. Therefore, the coefficient
$c_{2}$ must change its sign at the points $z_{i}$. Similar arguments can be used
to prove that the coefficient $c_{4}$ must change its sign at the points $z_{j}$
where $y(z_{j})=1$.

Thus $z$ changed
over half a period of $cn(z, {1\over \sqrt 2})$, a new decomposition of the solution
of eq. (8) should be made:
$$X=\tilde c_1 \phi_1 +\tilde c_2 \phi_2, \eqno 21$$
where in general the coefficients $\tilde c_{1,2}$ do not coincide with the
coefficients $c_{1,2}$. The rules for changing of these coefficients follow directly 
from the arguments
mentioned above if one uses eqs (9-15) taking
into account the explicit form of the connection coefficients
(16,17). It is straightforward to obtain the relation:
$$\tilde c^i =t^i_j c^j, \eqno 22$$
where the components of the matrix $t^i_j$ have the following explicit form:
$$t^1_1=t^2_2=\sqrt{2} \cos {\pi (\alpha -\beta)}, \eqno 23 $$
$$t^1_2={8\pi\Gamma^{2}(2-\gamma)\over \Gamma(1-\alpha)\Gamma(1-\beta)
\Gamma(1+\alpha -\gamma)\Gamma (1+\beta -\gamma )}, \eqno 24$$
$$t^2_1={8\pi\Gamma^{2}(\gamma)\over \Gamma(\alpha)\Gamma(\beta)
\Gamma(\gamma-\alpha)\Gamma (\gamma - \beta )}, \eqno 25$$
The eigenvalues of the matrix $t^i_j$ are
$$\lambda_{1,2}=\sqrt 2 \cos {\pi (\alpha -\beta)}\pm \sqrt{\cos{2\pi(\alpha -\beta)}},
 \eqno 26 $$ 
and $\alpha -\beta = {\sqrt{1+8\kappa^2}\over 4}$
\footnote{To obtain the eigenvalues $\lambda_{1,2}$, we use the well known relations
$\Gamma(x+1)=x\Gamma(x)$ and $\Gamma(1-x)\Gamma(x)={\pi \over \sin {(\pi x)}}$}.  

Obviously, the multiplicators $\rho_{1,2}$ are equal to
$\lambda_{1,2}^2$, and the characteristic exponents are 
$$\mu_{1,2} ={1\over T}\ln {\rho_{1,2}}={2\over T}\ln ({\sqrt 2 \cos {\pi (\alpha -\beta)}
\pm \sqrt{\cos{2\pi(\alpha -\beta)}} }). \eqno 27$$

In actual applications it is very important to know under which conditions a particular solution
to eq. (3) experiences parametric amplification, that is under which conditions
its amplitude increases when the ``time''
$z$ changes over the period of $cn(z,{1\over \sqrt 2})$. To characterize 
the parametric amplification, we introduce the  real quantity
$$\tilde \mu (\kappa )=Max (Re (\mu_{1,2})), \eqno 28$$
Obviously, $\tilde \mu (\kappa ) > 0$ is the condition for parametric 
amplification. It is easy to see that it is satisfied when $\lambda_{1,2}$ are real,
that is when
$$n-{1\over 4} < \alpha -\beta < n+{1\over 4}, \eqno 29$$
where n is an integer $\ge 1$. 
In terms of the parameter $\kappa $, these inequalities 
can be rewritten as
$$\sqrt {n(2n-1)} < \kappa < \sqrt {n(2n+1)}. \eqno 30$$
The expression for the quantity $\tilde \mu $ follows from eq. (27) and the definitions of
the parameters $\alpha $ and $\beta $ and the expression for the period
$T=4{\bf K}({1\over \sqrt 2})={\Gamma^{2}(1/4)\over \sqrt \pi}$:
$$\tilde \mu (\kappa )={2\sqrt \pi \over \Gamma^{2} ({1\over 4})}\ln {\lbrace 
{\sqrt 2|\cos {({\pi \sqrt{1+8\kappa^{2}}\over 4})}}|+
\sqrt{\cos{( {\pi\sqrt{1+8\kappa^{2}}\over 2})}}\rbrace } \eqno 31 $$
The quantity $\tilde \mu $ attains its maximal value at
$$\kappa =\sqrt {2n^2 -{1\over 8}}, \eqno 32$$
where its value is
$$\tilde \mu_{max}={2\sqrt \pi \over \Gamma^{2} ({1\over 4})}\ln {(1+\sqrt 2)}
\approx 0.2377. \eqno 33$$ 
Note that the same value has been obtained in [1] in the asymptotic
limit $\kappa \rightarrow \infty$.

\section{Discussion}

We were not able
to find the simple formulae derived in the standard reference books
on the Lame's equation [6-8]. However, a very similar transformation 
between another equation of Lame's type and a hyper-geometric equation
has been discussed recently by Clarkson and Olver [9] (see also [10]). 
They show that the hyper-geometric equation
$$t(1-t){d^2\over dt^2}\Psi +({1\over 2}-{7\over 6}t){d \over dt}\Psi-\sigma \Psi=0,
\eqno 34$$
and Lame's equation
$${d^2\over du^2}\Psi +36\sigma W(u)  \Psi =0, \eqno 35$$
where $W(u)$ is the Weierstrass elliptic function with parameters 
$g_{2}=0, g_{3}={1\over 3^{3/2}16}$,
are related to each other by the transformation 
$$cn(u, K)={\sqrt 3 -1 +{(1-t)}^{1/3}\over \sqrt 3 + 1 -{(1-t)}^{1/3}}, \eqno 36$$
where $K=\sqrt {{1\over 2}+{\sqrt 3\over 4}}$. 
Obviously, the characteristic exponents of equation (29) can be obtained
by a method similar to that described above. In general, it would be very
interesting to find a general solution to the following problem: under what condition
can Lame's equation of the general form be transformed to a  hyper-geometric equation?
The solution could be applied in many problems of modern
particle physics and cosmology.  

Finally, it is interesting to note that in principle the same line of argument could be
applied to Lame's equation of the general form. It is well known that after the change
of variable $y_{1}=cn^{2}(z,K)$  Lame's equation is reduced to a particular form of
the Heun's equation. Then the calculation of the characteristic exponents is reduced to
solving of the connection problem for Heun's equation between 
the elementary solutions corresponding to the singular points $y_{1}=0$ and $y_{1}=1$ 
(e.g. [11]. This problem can 
indeed be solved [11], but the connection coefficients analogous
to coefficients (16,17) are now expressed in terms of complicated series 
which look
rather difficult for analytic treatment of the general case. 

\section*{acknowledgments}

I am grateful to A. Illarionov, A. Kirillov and A. Starobinsky
for remarks. This work has been supported
in part by RFBR grant 
N 00-02-16135 and in part by the Danish Research Foundation through its establishment of
the Theoretical Astrophysics Center.

\section*{}
\Bibliography{9}
\item
[1]
Green P, Kofman L, Linde A and Starobinsky A 1997
{\it Phys. Rev. D} ${\bf 56}$
6175-6192
\item
[2] Dolgov A and Kirilova D 1990 {\it Sov. Nucl. Phys.} ${\bf 51}$ 273;
Traschen J and Brandenberger R 1990 {\it Phys. Rev. D} ${\bf 42}$ 2491;
Shtanov Y Traschen J and Brandenberger R 1995 {\it Phys. Rev. D}
${\bf 51}$
5438
\item
[3] Kofman L Linde A and Starobinsky A 1994
{\it Phys. Rev. Letters}
${\bf 73}$ 3195

\item
[4] Kofman L Linde A and Starobinsky A 1997 {\it Phys. Rev. D}
${\bf 56}$ 3258

\item
[5] Caratheodory C 1960 {\it Theory of Functions} v 2 (New York: Chelsea 
Publishing Company)
\item
[6] Whittaker E and Watson G 1996 {\it A Course of Modern Analysis}
(Cambridge: Cambridge University Press)
\item
[7] Bateman H and Erdelyi A 1955 
{\it Higher Transcendental Functions} v 3
(McGraw-Hill Book Company)
\item
[8] Arscott F 1964 {\it Periodic Differential Equations} 
(Pergamon Press)
\item
[9] Clarkson P and Olver P 1996 {\it J. of Differential Equations} {\bf 124} 225
\item
[10] Kamke E 1975 {\it Differentialgleichungen L$\ddot o$sungsmethoden and 
L$\ddot o$sungen} v 1
(Chelsea, New York) 
\item
[11] Schafke R and Schmidt D 1980 {\it SIAM J. Math. Anal.} ${\bf 11}$ 848

\endbib
\end{document}